\documentclass[conference, 9pt]{IEEEtran}

\ifCLASSINFOpdf

\else

\fi

\usepackage{amsmath,graphicx}
\usepackage{amssymb}

\usepackage{psfrag}
\usepackage[table,xcdraw,dvipsnames]{xcolor}
\usepackage{subfig}

\usepackage[T1]{fontenc}
\usepackage[font=small]{caption}

\DeclareCaptionLabelFormat{andtable}{#1~#2  \&  \tablename~\thetable}

\usepackage{pgfplots}
\pgfplotsset{compat=newest}
\pgfplotsset{plot coordinates/math parser=false} 
\usepgfplotslibrary{groupplots}

\usepackage{tikz}
\usetikzlibrary{matrix}
\newlength\figureheight
\newlength\figurewidth 

\usepackage{multirow}

\usepackage[customcolors]{hf-tikz}

\tikzset{style gray/.style={
    set fill color=gray!50,
    set border color=white,
  },
  hor/.style={
    above left offset={-0.15,0.31},
    below right offset={0.15,-0.125},
    #1
  },
  ver/.style={
    above left offset={-0.1,0.3},
    below right offset={0.15,-0.15},
    #1
  }
}

\begin{document}
%
\title{Graph-Based Compensated Wavelet Lifting for 3-D+t Medical CT Data}

\author{\IEEEauthorblockN{Daniela Lanz, Andr\'{e} Kaup}
\IEEEauthorblockA{Multimedia Communications and Signal Processing\\
Friedrich-Alexander-Universit\"at Erlangen-N\"urnberg (FAU), Cauerstr. 7, 91058 Erlangen, Germany\\
Email: \{Daniela.Lanz,Andre.Kaup\}@FAU.de\\}
}

\maketitle

\begin{abstract}
An efficient scalable data representation is an important task especially in the medical area, e.g. for volumes from Computed Tomography (CT) or Magnetic Resonance Tomography (MRT), when a downscaled version of the original signal is needed. Image and video coders based on wavelet transforms provide an adequate way to naturally achieve scalability. This paper presents a new approach for improving the visual quality of the lowpass band by using a novel graph-based method for motion compensation, which is an important step considering data compression. We compare different kinds of neighborhoods for graph construction and demonstrate that a higher amount of referenced nodes increases the quality of the lowpass band while the mean energy of the highpass band decreases. We show that for cardiac CT data the proposed method outperforms a traditional mesh-based approach of motion compensation by approximately~11~dB in terms of PSNR of the lowpass band. Also the mean energy of the highpass band decreases by around~30$\%$.
\end{abstract}

\IEEEpeerreviewmaketitle

\section{Introduction}
Multi-dimensional data volumes, like dynamical 3-D+t volumes, can get very large, which is challenging for accessing, transmitting and storing the data. Especially for telemedical applications, browsing and fast previewing are important tasks, where a downscaled version of the original signal is required. One possibility to reduce the filesize is to simply subsample a sequence by a factor of 2. However, not only the size of the data is reduced but also the remaining sequence of images is affected by severe temporal aliasing.

An adequate way to naturally achieve scalability features without additional overhead is based on subband coding~\cite{lnt2011-23}. A wavelet transform can overcome the drawback mentioned above by decomposing the signal in every transformation step into a lowpass and a highpass band with the energy concentrated in the lowpass band. Then the lowpass band serves as a representative for a particular pair of sequent frames, while the highpass band contains the structural information. Due to the motion of the CT or MRT volume the lowpass band can contain blur and ghosting artifacts. To overcome these artifacts caused by displacement in the signal, it is possible to incorporate a compensation method directly into the wavelet transform, so the transform can be adapted to the signal. A high visual quality of the lowpass band is very important when it shall be used as a scalable representation for medical applications. Additionally an adequate motion compensation method leads to a better energy compaction in fewer transform coefficients. The latter property can be exploited to gain a higher coding efficiency. Therefore in this paper we focus on the motion compensation, which is a key tool in the development of highly scalable video compression algorithms~\cite{958672}.

A compensated wavelet transform in temporal direction is known as Motion Compensated Temporal Filtering\linebreak (MCTF)~\cite{334985}. There exist several approaches for compensation like block-based and mesh-based compensation methods~\cite{lnt2012-40,lnt2014-23}. Beside the fact, that there are still artifacts remaining in the lowpass band, the required inversion of the motion compensation is not always as simple as assumed~\cite{lnt2013-18}. Interpreting images as graph signals allows us to overcome these drawbacks and incorporates the geometric structure of the data~\cite{shuman2013emerging}. In this paper we will present a novel approach of compensated wavelet lifting, which is based on signal processing on graphs.

Section~2 presents a brief review of the compensated wavelet lifting followed by a detailed description of image processing on graphs in Section~3. After defining the main terms of graph theory, a graph-based wavelet transform as mentioned in~\cite{narang2009lifting} is introduced. Then, the new graph-based motion compensation approach is presented. Simulation results are shown in Section~4 followed by a short conclusion in Section~5.

\section{Compensated Wavelet Lifting}
\label{sec:format}

\begin{figure}[tb]
  \begin{scriptsize}
  \centering
  \psfragscanon
  \psfrag{ref}{reference frame}
  \psfrag{cur}{current frame}
  \psfrag{x}{$x$}
  \psfrag{y}{$y$}
  \psfrag{time}{time}
  \psfrag{f1}{$f_{1}$}
  \psfrag{f2}{$f_{2}$}
  \psfrag{f1}{$f_{1}$}
  \psfrag{f2}{$f_{2}$}
  \psfrag{f3}{$f_{3}$}
  \psfrag{f4}{$f_{4}$}
  \psfrag{f2t-1}{$f_{2t-1}$}
  \psfrag{f2t}{$f_{2t}$}
  \psfrag{HP1}{$\text{HP}_1$}
  \psfrag{LP1}{$\text{LP}_1$}
  \psfrag{HP2}{$\text{HP}_2$}
  \psfrag{LP2}{$\text{LP}_2$}
  \psfrag{HPt}{$\text{HP}_t$}
  \psfrag{LPt}{$\text{LP}_t$}
  \psfrag{frac12}{$\frac1{2}$}
  \psfrag{MC}{MC}
  \psfrag{IMC}{IMC}
  \psfragscanoff
  \includegraphics[width=0.5\textwidth]{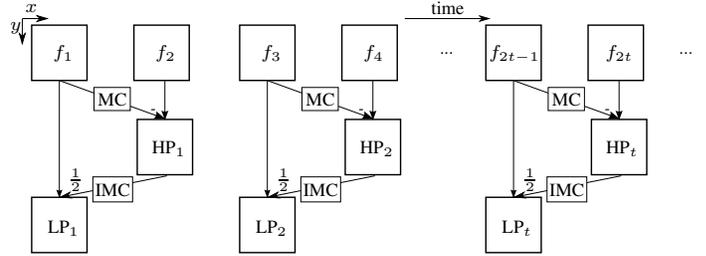}
  \caption{Compensated Haar lifting structure in temporal direction (MCTF).}
  \label{fig:lifting_scheme}
  \end{scriptsize}
\end{figure}

The lifting structure is a factorized representation of the wavelet transform, which allows the integration of arbitrary compensation methods into the transform~\cite{Sweldens1995}. Figure~\ref{fig:lifting_scheme} shows a schematic of the compensated lifting structure of the Haar wavelet. The frames $f_t$ are indexed by the time step $t$ in temporal direction. Analogously for spatial decomposition the frames $f_z$ get a slice index $z$. For the sake of simplicity, we will explain the following computations only for the temporal direction.

In the prediction step the highpass coefficients $\text{HP}_t$ are computed according to
\begin{equation}
\text{HP}_t = f_{2t}-\lfloor \mathcal{W}_{2t-1\rightarrow 2t}(f_{2t-1})\rfloor.
\end{equation}
To achieve a motion compensated transform, a predictor, denoted by the warping operator $\mathcal{W}_{2t-1\rightarrow 2t}$ is subtracted from the current frame $f_{2t}$, instead of the reference frame $f_{2t-1}$. The motion compensation is denoted by MC in Fig.~\ref{fig:lifting_scheme}. In the update step, the compensation has to be inverted to achieve an equivalent wavelet transform. This process is called inverse motion compensation (IMC). Then, the lowpass coefficients are computed by
\begin{equation}
\text{LP}_t = f_{2t-1}+\lfloor \frac{1}{2}\mathcal{W}_{2t\rightarrow 2t-1}(\text{HP}_t)\rfloor.
\end{equation}
To reconstruct the original sequence losslessly from the lowpass coefficients, floor operators are applied in the lifting structure to avoid rounding errors~\cite{647983}. This is a very important aspect considering medical data, like CT or MRT volumes.

\section{Image Processing on Graphs}
\label{sec:graphbasedIP}
Considering images as graph signals, images can be represented as undirected graphs $G(\mathcal V,E)$, where $\mathcal{V}$ is the set of pixels, which are corresponding to nodes, indexed as $1,2,3,...,N$, and $E$ is the set of links between nodes. Every link is defined by a triplet $(i,j,w_{ij}) $, where $i$ and $j$ are the start and end nodes respectively and $w_{ij}$ is the weight which is often measured as the spatial or photometric similarity between the nodes $i$ and $j$ \cite{6694319}. If $i$ and $j$ are linked to each other, this will be denoted as $i \sim j$. There are various ways to connect graphs, like 4-adjacency grid graphs, 8-adjacency grid graphs or 25~nearest-neighbors graphs. $G$ is further characterized by its weighted adjacency matrix $\mathbf{W}$ given by  
\begin{equation}
w_{i,j} = \begin{cases}
     w_{ij} & \text{if } i \sim j  \\
     0 & \text{otherwise.} 
   \end{cases}
\end{equation}
In case of an unweighted graph, the adjacency matrix is denoted as $\mathbf{A}$ and contains a $1$ if $i \sim j$ and a $0$ otherwise. The value of each node is given by the vector $X$, such that $X(i)$ is the intensity $f_t(i)$ of pixel $i$ in the considered frame at time step $t$ \cite{hidane2013lifting}. Keeping these definitions in mind, we will introduce lifting-based wavelet transforms on graphs. 

\subsection{Graph-Based Wavelet Transform}
\label{subsec:graphbasedWT}
\begin{figure}[tb]
  \centering
  \psfragscanon
  \psfrag{X}{$X = \begin{pmatrix}
  X_\text{even} \\ 
  X_\text{odd}
  \end{pmatrix}$}
  \psfrag{H}{$H$}
  \psfrag{L}{$L$}
  \psfrag{K}{$\mathbf{K_U}$}
  \psfrag{J}{$\mathbf{J_P}$}
  \psfragscanoff
  \includegraphics[width=0.45\textwidth]{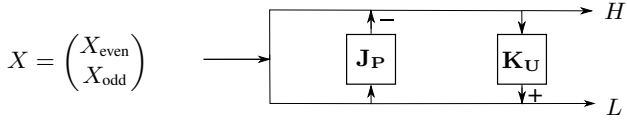}
  \caption{Block diagram for a lifting transform.}
  \label{fig:graphbased_lifting}
\end{figure}
In order to apply a lifting-based transform to an arbitrary graph, the set $\mathcal{V}$ of nodes has to be split into two disjoint subsets of even-odd assignment, so that every node of one subset has only neighbors which belong to the other subset~\cite{narang2009lifting}. By rearranging the vector $X$ of nodes to gather  $l$ odd and $m=N-l$ even nodes at one place and rearrange the adjacency matrix accordingly, we get a representation
\begin{equation}
X = \begin{pmatrix}
X_\text{even} \\ 
X_\text{odd}
\end{pmatrix} \quad 
\mathbf{A} = \begin{pmatrix}
\mathbf{F}_{m\times m} & \mathbf{J}_{m\times l} \\ \mathbf{K}_{l\times m} & \mathbf{L}_{l\times l}
\end{pmatrix}
\end{equation}
where $X_\text{odd}$ is a $l \times 1$ vector and $X_\text{even}$ is a $m\times 1$ vector. The submatrices $\mathbf{F}$ and $\mathbf{L}$ contain adjacencies of only odd and even nodes respectively. So these matrices contain links which have conflicts since they connect nodes of same parity. In contrast the submatrices $\mathbf{J}$ and $\mathbf{K}$ contain links without conflicts. In case of perfect splitting, the matrices $\mathbf{F}$ and $\mathbf{L}$ remain empty.

Once the splitting of nodes in the graph is done, a lifting-based wavelet transform can be applied:
\begin{equation}
\begin{aligned}
H &= X_\text{even}-\mathbf{J_P}\times X_\text{odd}\\
L &= X_\text{odd}+\mathbf{K_U}\times H.
\end{aligned}
\end{equation}
Matrix $\mathbf{J_P}$ describes the prediction matrix computed from matrix $\mathbf{J}$ and leads to the required MC when multiplying it by $X_\text{odd}$. Similarly, $\mathbf{K_U}$ stands for the update matrix, computed from matrix $\mathbf{K}$. Multiplying $\mathbf{K_U}$ by $H$ inverses the MC. Vector $H$ contains the highpass coefficients whereas vector $L$ contains the lowpass coefficients. In Fig.~\ref{fig:graphbased_lifting}, the schematic process of the computations above can be seen. 

Since this transform is invertible, the original values can be recovered by applying the following inverse lifting steps:
\begin{equation}
\begin{aligned}
X_\text{even} &= L-\mathbf{K_U}\times H\\
X_\text{odd} &= H+\mathbf{J_P}\times X_{even}.
\end{aligned}
\end{equation}
The design of the matrices $\mathbf{J_P}$ and $\mathbf{K_U}$ depends on the application for which the wavelet transform is applied, for example denoising, smoothing, or filtering of the given graph. So the prediction and update weights have to be adapted to the desired application.

\subsection{Proposed Graph-Based Motion Compensation}
\label{subsec:graphbasedMC}
\begin{figure}[tb]
  \centering
  \psfragscanon
  \psfrag{odd}{Odd set of nodes}
  \psfrag{even}{Even set of nodes}
  \psfrag{ref}{$f_{2t-1}$}
  \psfrag{cur}{$f_{2t}$}
  \psfragscanoff
  \includegraphics[trim=0 0 0 0,width=0.35\textwidth]{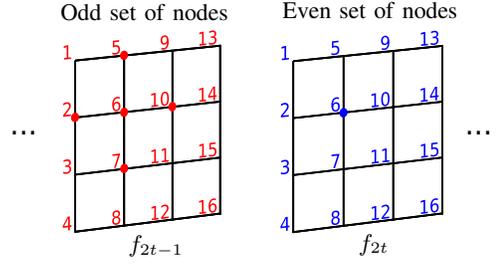}
  \caption{Node assignment to build the corresponding adjacency matrices.}
  \label{fig:adj_splitting}
\end{figure}

The novelty of this work is to consider a whole sequence of images as a \mbox{3-D} graph signal instead of denoting every single frame as a \mbox{2-D} graph. This sounds very simple, but solves the problem of a proper partitioning of the nodes mentioned in \cite{hidane2013lifting}. Then the required splitting of nodes can easily be reached by assigning every frame according to its number of appearance as even and odd as shown in Fig.~\ref{fig:adj_splitting}. Here the red frame contains the set of odd nodes, while the blue frame contains the set of even nodes. Within the given frames, the given numbers explain the order of the nodes, in which they are getting rearranged in the corresponding vectors $X_\text{odd}$ and $X_\text{even}$. For the sake of formality we further name these vectors $X_\text{ref}$ and $X_\text{cur}$ respectively, according to the previously used notation of reference and current frames.

\begin{figure}[b]
  \centering
  \psfragscanon
  \psfrag{ref}{\parbox{4 cm}{Reference frame $f_{2t-1}$}}
  \psfrag{cur}{\parbox{4 cm}{Current frame $f_{2t}$}}
  \psfrag{8}{\parbox{4cm}{8-grid\\neighborhood}}
  \psfrag{4}{\parbox{4cm}{4-grid\\neighborhood}}
  \psfrag{25}{\parbox{4cm}{25-nearest\\neighborhood}}
  \psfragscanoff
  \includegraphics[clip=60 0 0 49,width=0.36\textwidth]{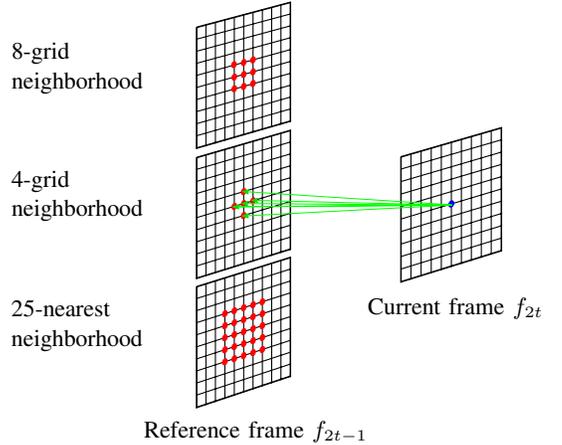}
  \caption{Possible neighborhoods for one single node of the current frame connected to the reference frame using the novel approach of a 3-D graph signal.}
  \label{fig:sequence_splitting}
\end{figure}

By doing this, a perfect splitting can be reached, so the submatrices $\mathbf{F}$ and $\mathbf{L}$ remain empty. Further, the submatrices $\mathbf{J}$ and $\mathbf{K}$ are described by the desired neighborhood for every single node in the reference frame $f_{2t-1}$ related to the current frame $f_{2t}$.

To clarify this new way of linking even and odd nodes adequately, Fig.~\ref{fig:sequence_splitting} shows how a single node of the current frame $f_{2t}$, represented as blue dot, is connected according to the chosen neighborhood to its corresponding nodes in the reference frame $f_{2t-1}$, which are illustrated as red dots. The figure shows the connectivity for a 4-grid, 8-grid and 25-nearest neighborhood. This process has to be done for every single node of the current frame $f_{2t}$ to the reference frame $f_{2t-1}$ to obtain $\mathbf{J}$ and vice versa to get $\mathbf{K}$.

In the next step, the selected neighborhood between every two frames is stored in the matrices $\mathbf{J}$ and $\mathbf{K}$. Equation~(\ref{eq:adj}) gives an example for $\mathbf{J}$ for a 4-grid neighborhood. The gray colored row shows node $6$ of the current frame $f_{2t}$, represented as a blue dot in Fig. \ref{fig:adj_splitting}, and its neighboring nodes $2,5,6,7 \text{ and } 10$ in the reference frame $f_{2t-1}$, represented as red dots.
\begin{equation}\resizebox{0.81\hsize}{!}{
{\large \bf{J} =} \bordermatrix{
  & \color{red}1 & \color{red}2 & \color{red}3 & \color{red}4 & \color{red}5 & \color{red}6 & \color{red}7 & \color{red}8 & \color{red}9 & \color{red}10& \color{red}11& \color{red}12& \color{red}13& \color{red}14& \color{red}15 & \color{red}16\cr
\color{blue}1 & 1 & 1 & 0 & 0 & 1 & 0 & 0 & 0 & 0 & 0 & 0 & 0 & 0 & 0 & 0 & 0\cr
\color{blue}2 & 1 & 1 & 1 & 0 & 0 & 1 & 0 & 0 & 0 & 0 & 0 & 0 & 0 & 0 & 0 & 0\cr
\color{blue}3 & 0 & 1 & 1 & 1 & 0 & 0 & 1 & 0 & 0 & 0 & 0 & 0 & 0 & 0 & 0 & 0\cr
\color{blue}4 & 0 & 0 & 1 & 1 & 0 & 0 & 0 & 1 & 0 & 0 & 0 & 0 & 0 & 0 & 0 & 0\cr
\color{blue}5 & 1 & 0 & 0 & 0 & 1 & 1 & 0 & 0 & 1 & 0 & 0 & 0 & 0 & 0 & 0 & 0\cr
\color{blue}6 & \tikzmarkin[hor=style gray]{el} 0 & 1 & 0 & 0 & 1 & 1 & 1 & 0 & 0 & 1 & 0 & 0 & 0 & 0 & 0 & 0\tikzmarkend{el}\cr
\color{blue}7 & 0 & 0 & 1 & 0 & 0 & 1 & 1 & 1 & 0 & 0 & 1 & 0 & 0 & 0 & 0 & 0\cr
\color{blue}8 & 0 & 0 & 0 & 1 & 0 & 0 & 1 & 1 & 0 & 0 & 0 & 1 & 0 & 0 & 0 & 0\cr
\color{blue}9 & 0 & 0 & 0 & 0 & 1 & 0 & 0 & 0 & 1 & 1 & 0 & 0 & 1 & 0 & 0 & 0\cr
\color{blue}10& 0 & 0 & 0 & 0 & 0 & 1 & 0 & 0 & 1 & 1 & 1 & 0 & 0 & 1 & 0 & 0\cr
\color{blue}11& 0 & 0 & 0 & 0 & 0 & 0 & 1 & 0 & 0 & 0 & 1 & 1 & 0 & 0 & 1 & 0\cr
\color{blue}12& 0 & 0 & 0 & 0 & 0 & 0 & 0 & 1 & 0 & 0 & 1 & 1 & 0 & 0 & 0 & 1\cr
\color{blue}13& 0 & 0 & 0 & 0 & 0 & 0 & 0 & 0 & 1 & 0 & 0 & 0 & 1 & 1 & 0 & 0\cr
\color{blue}14& 0 & 0 & 0 & 0 & 0 & 0 & 0 & 0 & 0 & 1 & 0 & 0 & 1 & 1 & 1 & 0\cr
\color{blue}15& 0 & 0 & 0 & 0 & 0 & 0 & 0 & 0 & 0 & 0 & 1 & 0 & 0 & 1 & 1 & 1\cr
\color{blue}16& 0 & 0 & 0 & 0 & 0 & 0 & 0 & 0 & 0 & 0 & 0 & 1 & 0 & 0 & 1 & 1\cr
}}
\label{eq:adj}
\end{equation}

Now, the matrices $\mathbf{J_P}$ and $\mathbf{K_U}$ have to be computed to reach the required MC and IMC. In this novel method the computation is done by three steps:
\begin{enumerate}
\item At first the matrices $\mathbf{J}$ and $\mathbf{K}$ are weighted in accordance to the constraints mentioned at the beginning of this section. According to \cite{Bougleux2005} the weight function
\begin{equation}
w_{ij} = e^{-\lambda\lvert f_{2t-1}(j)-f_{2t}(i)\rvert}, \lambda \in \mathbb{R}_+
\end{equation}
estimates the similarity between two pixels. In this work we chose $\lambda=0.5$.
\item Further the diagonal matrices $\mathbf{D}$ of the weighted matrices $\mathbf{W_J}$ and $\mathbf{W_K}$ are computed respectively by
\begin{equation}
d_{ii} = \sum_j \bf{W_{ij}}.
\end{equation}
\item
Using random walks on Graph $G$ delivers the Markov chain with transition matrix 
\begin{equation}
\bf{P = D^{-1}W}.
\end{equation}
According to \cite{lee2011multiscale}, $p_{ij}$ is the probability of being at node $j$ starting from node $i$. Finally the matrices $\mathbf{J_P}$ and $\mathbf{K_U}$ are defined by
\begin{equation}
\begin{aligned}
\bf{J_P} &= \bf{P_J}\\
\bf{K_U} &= \bf{P_K}.
\end{aligned}
\end{equation}
\end{enumerate}
The sparsity of these matrices is one of the most important conditions considering the required coding as the next step in the process chain of image and video compression.
 
Using these specifications in the context of the graph-based wavelet transform between every pair of frames of a sequence and keep the formulas of the Haar wavelet in mind, we get the transform
\begin{equation}
\begin{aligned}
H &= X_\text{cur}-\lfloor\mathbf{J_P}\times X_\text{ref}\rfloor\\
L &= X_\text{ref}+\lfloor\frac{1}{2}\mathbf{K_U}\times H\rfloor
\end{aligned}
\end{equation}
and the corresponding inverse transform
\begin{equation}
\begin{aligned}
X_\text{ref} &= L-\lfloor\frac{1}{2}\mathbf{K_U}\times H\rfloor\\
X_\text{cur} &= H+\lfloor\mathbf{J_P}\times X_\text{ref}\rfloor.
\end{aligned}
\end{equation}
The term $\mathbf{J_P}\times X_\text{ref}$ describes an estimation for $X_\text{cur}$, whereas $\mathbf{K_U}\times H$ reverses the compensation in the update step. This inversion is very easy to compute and provides good results. By rearranging the vectors $H$ and $L$ after the transform step and accordingly $X_\text{cur}$ and $X_\text{ref}$ after the inverse transform step in reversed order, the highpass and lowpass band HP and LP or rather the current and reference frames are achieved again.

Since the proposed approach is a non iterative method, the computational complexity is low compared to the mesh-based approach. 

\section{Simulation Results}
\label{sec:results}
For the evaluation of the proposed method, we used a multidimensional \textit{cardiac} 3-D+t CT data set\footnote{The CT volume data set was kindly provided by Siemens Healthcare.}, which has 10 time steps, each with 127 slices in $z$-direction, and a resolution of $512\times 512$ pixels in $xy$-direction at 12 bit per sample and describes a beating heart over time.

In our simulation we perform one Haar wavelet decomposition step with and without various compensation methods in $z$-direction and in $t$-direction respectively. At first, we show the performance of the proposed method on two adjacent slices in $z$- and $t$- direction respectively. To evaluate the visual quality of the computed lowpass bands, we calculate the PSNR. Since it's not defined, whether the lowpass band should be a representative for the image $f_1 $ or $f_2$, we will calculate the PSNR for both cases, i.e. PSNR$(f_1,LP1)$ and PSNR$(f_2,LP1)$. Table~\ref{tab:res} provides the results for a 4-grid, 8-grid and 25-nearest neighborhood. The first row gives a reference value for the similarity of the underlying original frames denoted as PSNR$(f_1,f_2)$. The following two rows indicate the quality of the lowpass band compared to the reference frame and the current frame respectively. As can be seen, the lowpass band is more similar to the reference frame $f_1$ than to the current frame $f_2$, so further we assume the lowpass band as a representative for the reference frame $f_1$. The table proves that it is possible to achieve a very high visual quality for the lowpass band by using the novel graph-based motion compensation both in spatial as well as in temporal direction. On closer examination of Table~\ref{tab:res} the conclusion arrives, that a higher number of neighboring nodes leads to a better visual quality of the lowpass band compared to the reference frame $f_1$.

\begin{table}[t]
\centering
\begin{tabular}{|cc|c|c|}
\hline
\rowcolor[HTML]{C0C0C0} 
\multicolumn{1}{|c|}{\cellcolor[HTML]{C0C0C0}} & 4-grid & 8-grid & 25-nearest \\ \cline{2-4} 
\multicolumn{1}{|c|}{\multirow{-2}{*}{\cellcolor[HTML]{C0C0C0}\begin{tabular}[c]{@{}c@{}}PSNR $(f_x,f_y)$\\ {[}dB{]}\end{tabular}}} & \multicolumn{3}{c|}{Spatial decomposition} \\ \hline
\multicolumn{1}{|c|}{$(f_1,f_2)$} & 35.03 & 35.03 & 35.03 \\ \hline
\multicolumn{1}{|c|}{$(f_1,LP1)$} & 45.88 & 48.22 & 56.99 \\ \hline
\multicolumn{1}{|c|}{$(f2,LP1)$} & 37.23 & 36.46 & 35.27 \\ \hline
\multicolumn{1}{|l}{} & \multicolumn{3}{c|}{Temporal decomposition} \\ \hline
\multicolumn{1}{|c|}{$(f_1,f_2)$} & 37.32 & 37.32 & 37.32 \\ \hline
\multicolumn{1}{|c|}{$(f_1,LP1)$} & 48.91 & 51.22 & 58.46 \\ \hline
\multicolumn{1}{|c|}{$(f_2,LP1)$} & 39.23 & 38.51 & 37.62 \\ \hline
\end{tabular}
\caption{Results for one decomposition step in spatial and temporal direction.}
\label{tab:res}
\end{table}

\begin{figure}[tb]
  \centering
  \psfragscanon
  \psfrag{4}{$r=4$}
  \psfrag{2}{$r=2$}
  \psfragscanoff
  \includegraphics[clip=0 0 0 0,width=0.2\textwidth]{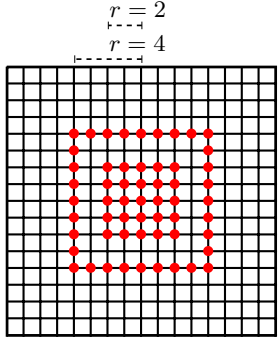}
  \caption{Increasing radius of neighboring nodes.}
  \label{fig:radius}
\end{figure}

To verify this assumption, we measure the quality for the lowpass band for an increasing radius of nodes taking into account as shown in Fig.~\ref{fig:radius} for computing the motion compensated frame in the prediction step. The results are shown in Table~\ref{tab:rad}, where $r=2$ corresponds to 25-nearest neighbors. Actually the PSNR$(f_1,LP1)$ increases with every level of a bigger radius and leads to a better visual quality with a higher energy compaction in the lowpass band in spatial as well as in temporal direction. 

\begin{table}[tb]
\centering
\begin{tabular}{c|c|c|}
\cline{2-3}
 & \multicolumn{2}{c|}{\cellcolor[HTML]{C0C0C0}PSNR$(f_1,LP1)${[}dB{]}} \\ \hline
\multicolumn{1}{|c|}{\cellcolor[HTML]{C0C0C0}Radius r} & Spatial & 
Temporal \\ \hline
\multicolumn{1}{|c|}{2} & 56.99 & 58.46 \\ \hline
\multicolumn{1}{|c|}{3} & 64.28 & 64.46 \\ \hline
\multicolumn{1}{|c|}{4} & 69.15 & 70.03 \\ \hline
\multicolumn{1}{|c|}{5} & 72.41 & 74.63 \\ \hline
\multicolumn{1}{|c|}{6} & 74.78 & 77.70 \\ \hline
\multicolumn{1}{|c|}{7} & 76.95 & 79.91 \\ \hline
\end{tabular}
\caption{Visual quality in terms of PSNR for the lowpass band $LP1$ compared to $f_1$ for different radii of neighboring nodes both in spatial and in temporal direction.}
\label{tab:rad}
\end{table}

\begin{figure}[htb]
\centering 
	 {\input{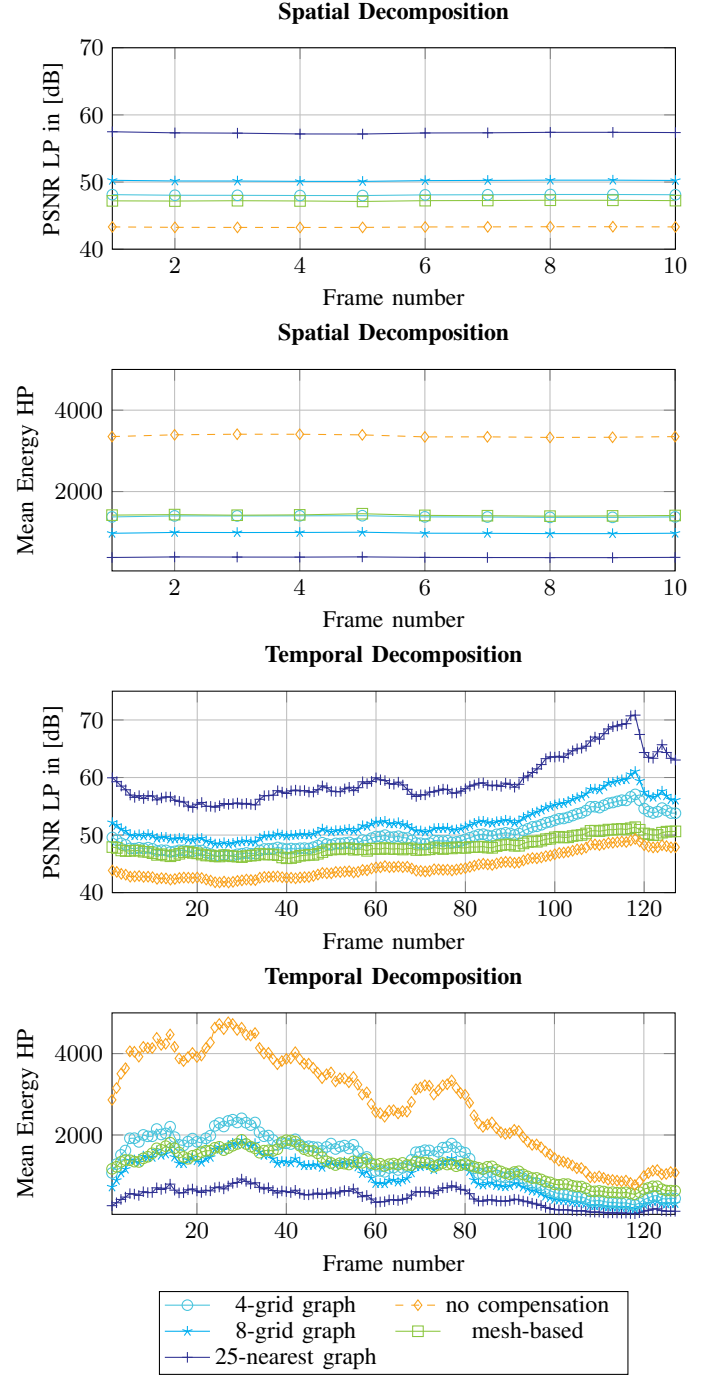}}
\caption{The quality of the lowpass band in terms of PSNR($f_{t,z}$,LP) in dB and the mean squared energy of the highpass band HP for no compensation, mesh-based, and graph-based compensation for both spatial as well as temporal decomposition.}
\label{fig:all} 
\end{figure}

Further, we analyze the results not only for one pair of frames, but for the whole volume, both in spatial and temporal direction for a 4-grid, 8-grid and 25-nearest neighborhood. We compare it to a mesh-based compensation and a simple Haar wavelet transform without compensation in spatial and in temporal direction. According to~\cite{lnt2012-29}, we used a quadrilateral mesh with a grid size of 16 $\times$ 16 pixels for the mesh-based approach. Figure~\ref{fig:all} shows the quality of the lowpass band and the mean energy of the highpass band for each method respectively. The averaged values can be seen in Table~\ref{tab:average}.

As expected, the PSNR for all methods including a compensation for both decomposition directions is significantly higher than the wavelet transform without a compensation method. Analogously the mean energy of the highpass band, which can be regarded as the prediction error for a compensated wavelet transform, lies significantly under the dashed line for all other methods, which represents the wavelet transform without compensation.

However, Fig.~\ref{fig:all} proves, that the graph-based compensation method outperforms the mesh-based approach not only in terms of visual quality of the lowpass band but also in case of the mean energy of the highpass band. While a 4-grid neighborhood provides already an average gain of 0,86 dB in spatial and 1,89 dB in temporal direction compared to the mesh-based approach, this can be further increased by using a 8-grid neighborhood which results in a gain of 2,98 dB and 4,31 dB respectively. The best results in this simulation can be reached by applying a 25-nearest neighborhood with an average gain of 10,11 dB and 11,53 dB in spatial and temporal direction respectively. For this case, also the mean energy of the highpass band reaches the best results. For the spatial decomposition it averages out at 27$\%$ compared to the mesh-based compensation and 37$\%$ for the temporal decomposition. By using other neighborhoods with increasing radius as introduced above, a further improvement can be expected.

\begin{figure}[htb]
\captionsetup[subfloat]{farskip=1pt,captionskip=1pt}
\centering 
     \subfloat[][reference frame $f_{2t-1}$]{\includegraphics[width=0.165\textwidth]{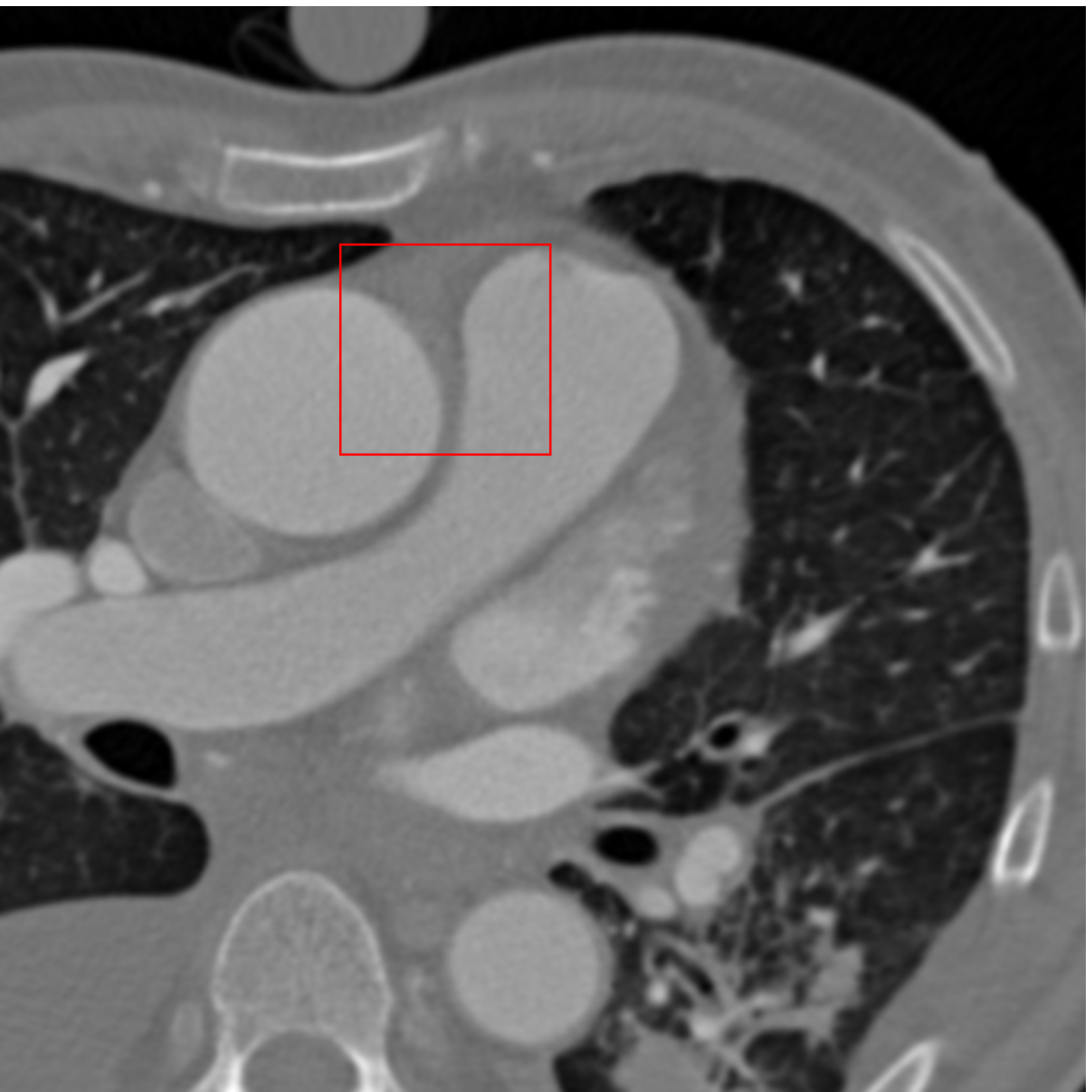}\label{fig:ref}} \quad
     \subfloat[][current frame $f_{2t}$]{\includegraphics[width=0.165\textwidth]{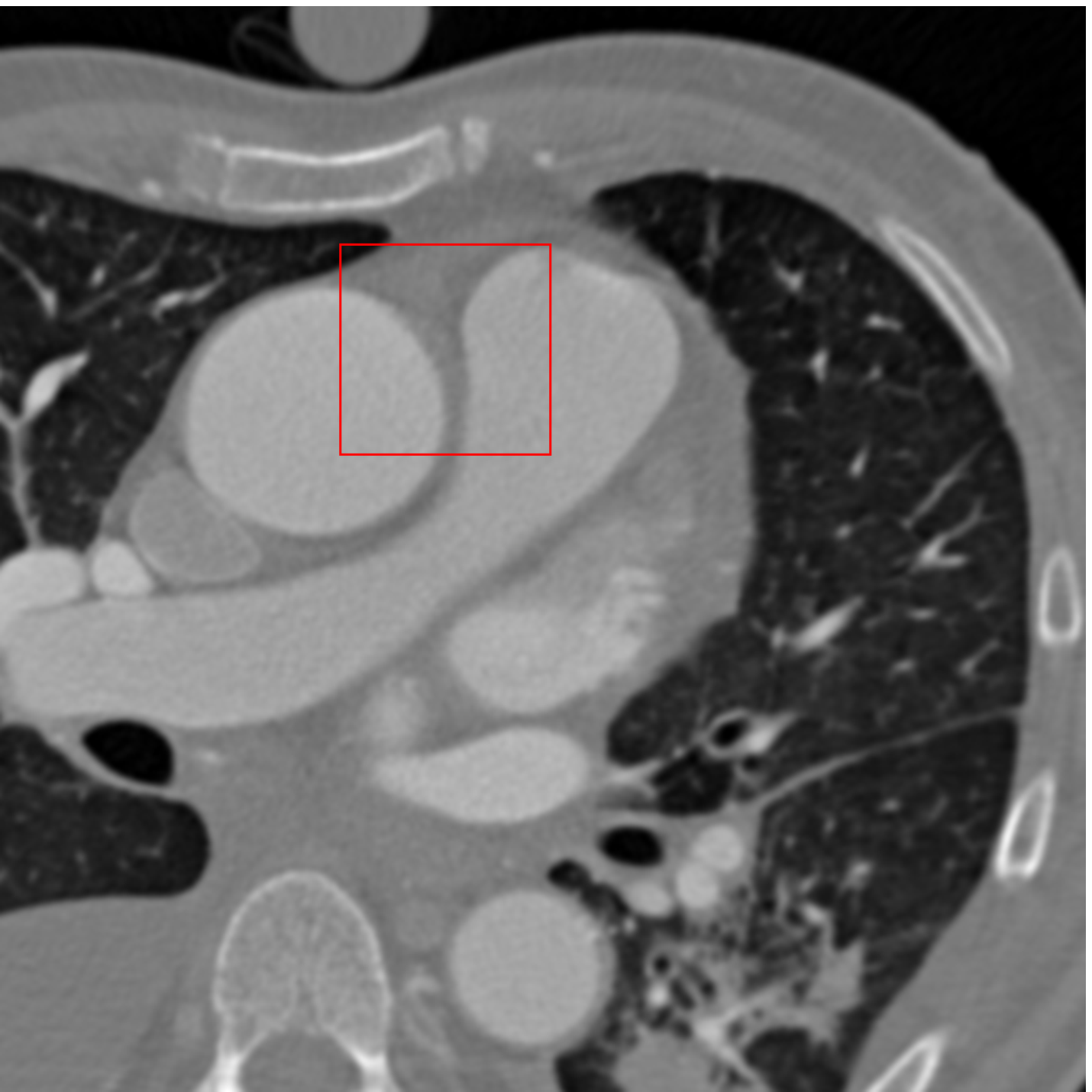}\label{fig:cur}}
     \hfill
     \subfloat[][no compensation LP]{\includegraphics[width=0.165\textwidth]{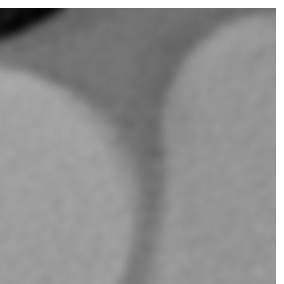}\label{fig:none_LP}} \quad
      \subfloat[][no compensation HP]{\includegraphics[width=0.165\textwidth]{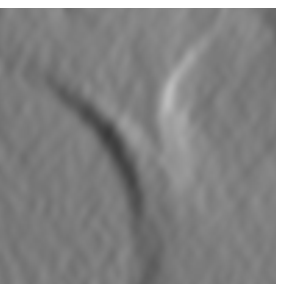}\label{fig:none_HP}} \hfill
     \subfloat[][mesh-based LP]{\includegraphics[width=0.165\textwidth]{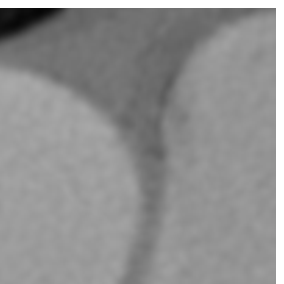}\label{fig:mesh_LP}}
     \quad
     \subfloat[][mesh-based HP]{\includegraphics[width=0.165\textwidth]{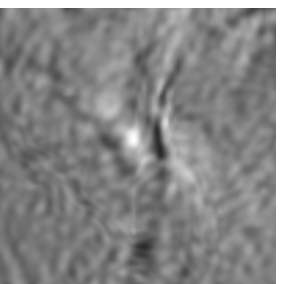}\label{fig:mesh_HP}}\hfill
     \subfloat[][25-nearest graph LP]{\includegraphics[width=0.165\textwidth]{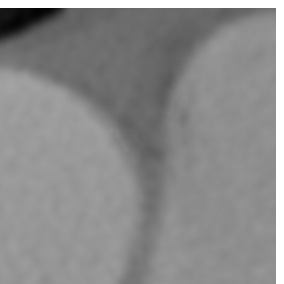}\label{fig:graph_LP}}
     \quad
     \subfloat[][25-nearest graph HP]{\includegraphics[width=0.165\textwidth]{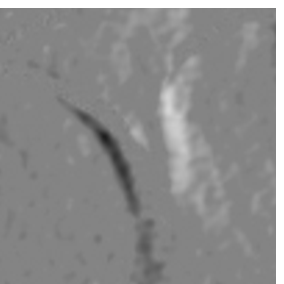}\label{fig:graph_HP}}
\caption{The first row shows temporal subsequent images of the original volume at one slice position. The red areas mark the origin of the details depicted in the remaining rows. Details from the highpass bands (HP) are shown in the right column and details from the lowpass bands (LP) are shown in the left column. The respective compensation methods are listed below. 
} 
\label{fig:compare}   
\end{figure}

The visual examples are depicted in Fig.~\ref{fig:compare} and illustrate the results from Table~\ref{tab:average}. The blurryness of the lowpass band without compensation can be sharpened by applying compensation methods. Using a graph-based compensation leads to a further reduction of the energy in the highpass band resulting in a lowpass band with less artifacts.

\begin{table}[t]
\centering
\begin{tabular}{|cc|c|}
\hline
\multicolumn{1}{|c|}{\cellcolor[HTML]{C0C0C0}} & \cellcolor[HTML]{C0C0C0} & \cellcolor[HTML]{C0C0C0} \\
\multicolumn{1}{|c|}{\cellcolor[HTML]{C0C0C0}} & \multirow{-2}{*}{\cellcolor[HTML]{C0C0C0}\begin{tabular}[c]{@{}c@{}}PSNR $LP$\\ {[}dB{]}\end{tabular}} & \multirow{-2}{*}{\cellcolor[HTML]{C0C0C0}\begin{tabular}[c]{@{}c@{}}Mean Energy\\ $HP$\end{tabular}} \\ \cline{2-3} 
\multicolumn{1}{|c|}{\multirow{-3}{*}{\cellcolor[HTML]{C0C0C0}Compensation Method}} & \multicolumn{2}{c|}{Spatial} \\ \hline
\multicolumn{1}{|c|}{none} & 43.30 & 3365.67 \\ \hline
\multicolumn{1}{|c|}{mesh-based} & 47.22 & 1418.67 \\ \hline
\multicolumn{1}{|c|}{4-grid graph} & 48.08 & 1385.25 \\ \hline
\multicolumn{1}{|c|}{8-grid graph} & 50.20 & 981.90 \\ \hline
\multicolumn{1}{|c|}{25-nearest graph} & 57.32 & 384.67 \\ \hline
\multicolumn{1}{|c|}{$\Delta$: 25-nearest graph to mesh-based} & +10.11 & -1034.00 \\ \hline
 & \multicolumn{2}{c|}{Temporal} \\ \hline
\multicolumn{1}{|c|}{none} & 44.57 & 2832.83 \\ \hline
\multicolumn{1}{|c|}{mesh-based} & 48.00 & 1236.68 \\ \hline
\multicolumn{1}{|c|}{4-grid graph} & 49.89 & 1343.13 \\ \hline
\multicolumn{1}{|c|}{8-grid graph} & 52.31 & 1006.44 \\ \hline
\multicolumn{1}{|c|}{25-nearest graph} & 59.53 & 459.20 \\ \hline
\multicolumn{1}{|c|}{$\Delta$: 25-nearest graph to mesh-based} & +11.53 & -777.50 \\ \hline
\end{tabular}
\caption{The table lists averaged results for the considered compensation methods. The last row provides a delta between the mesh-based and the 25-nearest graph compensation.}
\label{tab:average}
\end{table}

\section{Conclusion}
\label{sec:conclusion}

In this paper we introduced a novel graph-based compensated wavelet transform for medical 3-D+t volumes that contain deforming displacements over time. To overcome the problem of a proper partitioning of nodes required for a graph-based wavelet transform, we take a whole video sequence as a 3-D graph signal. The proposed method is able to provide a high quality lowpass band and a highpass band with a low mean energy. Other compensation methods may also provide good results for the motion compensated frame in the prediction step, but the inversion in the update step leads to erroneous results. While the block-based approach involves annoying artifacts in the lowpass band because of unconnected pixels, the mesh-based approach accepts an error by using only an approximation term instead of the quite complex inversion in the update step.

Further investigations aim at using segmented adjacency matrices to make sure that only nodes that refer to the same texture are getting connected by links. Also appropriate ways to encode the adjacency matrices should be found to make the graph-based compensated wavelet transform feasible for data compression.

\section*{Acknowledgment}
We gratefully acknowledge that this work has been supported by the Deutsche Forschungsgemeinschaft (DFG) under contract number KA 926/4-3.
\vspace{-2.1pt}
\bibliographystyle{IEEEtran}
\bibliography{Literatur1}

\end{document}